# Pseudo-PFLOW: Development of nationwide synthetic open dataset for people movement based on limited travel survey and open statistical data

Takehiro Kashiyama[a,1], Yanbo Pang[b], Yoshihide Sekimoto[b] and Takahiro Yabe[c]

[a]Institute of Industrial Science, University of Tokyo, Ce507, 4-6-1, Komaba, Meguro-Ku, Tokyo 153-8505, Japan, ksym@iis.u-tokyo.ac.jp
[b]Center for Spatial Information Science, The University of Tokyo, Ce507, 4-6-1, Komaba, Meguro-Ku, Tokyo 153-8505, Japan, [pybdtc, Sekimoto] @csis.u-tokyo.ac.jp
[c]Institute for Data, Systems, and Society, Massachusetts Institute of Technology, 50 Ames St, Cambridge, MA 02142, USA, tyabe@mit.edu

Correspondence
Takehiro Kashiyama, Institute of Industrial Science, University of Tokyo, Ce507, 4-6-1, Komaba, Meguro-Ku, Tokyo 153-8505, Japan, ksym@iis.u-tokyo.ac.jp

Present address
[1]Faculty of Economics, Osaka University of Economics, J611, 2-2-8, Osumi, Higashiyodogawa-Ku, Osaka 533-8533, Japan, t.kashiyama@osaka-ue.ac.jp

# Highlights

- A method for generating pseudo-people-flow data is proposed.
- Pseudo-people-flow data represent typical daily movement.
- The dataset for the entire population of Japan is published for researchers.
- The dataset is evaluated using mobile phone and travel survey data.

# Abstract

People flow data are utilized in diverse fields such as urban and commercial planning and disaster management. However, people flow data collected from mobile phones, such as using global positioning system and call detail records data, are difficult to obtain because of privacy issues. Even if the data were obtained, they would be difficult to handle. This study developed pseudo-people-flow data covering all of Japan by combining public statistical and travel survey data from limited urban areas. This dataset is not a representation of actual travel movements but of typical weekday movements of people. Therefore it is expected to be useful for various purposes. Additionally, the dataset represents the seamless movement of people throughout Japan, with no restrictions on coverage, unlike the travel surveys. In this paper, we propose a method for generating pseudo-people-flow and describe the development of a "Pseudo-PFLOW" dataset covering the entire population of approximately 130 million people. We then evaluated the accuracy of the dataset using mobile phone and trip survey data from multiple metropolitan areas. The results showed that a coefficient of determination of more than 0.5 was confirmed for comparisons regarding population distribution and trip volume.

# Keywords

People movement, Travel demand, Human mobility, Traffic simulation, Open data

# 1. Introduction

The necessity to gather people's movement data is high as they are used in various fields such as urban planning and commercial strategies. In recent years, research on the analysis of human movement data acquired from mobile phones, such as using global positioning system (GPS) and base station logs installed on smartphones, has attracted much attention. However, the raw data obtained from mobile phones are not provided externally by mobile carriers to protect privacy of users. Even if they are available, their handling is costly because they must be managed with a high level of security awareness. Recently, purchasing anonymized mobile phone data has become possible by aggregating them into trip volume data or grid population volume data, but this is highly expensive. Thus, mobile phone data are generally difficult to obtain, except for within limited cities or limited sample rates.

However, research has been conducted to construct various types of models of people's movement using mobile phone data. The most common method is to estimate the amount of movement between regions by determining model coefficients from data based on a spatial interaction model such as a gravity model or a radiation model [1–4]. Other approaches, such as motif analysis, have been used to extract people's movement patterns [5–9]. Recently, research has been conducted to construct behavioral models using machine learning and generate human flow data based on these models [10–13]. However, studies on spatial interaction models and motif analysis can reproduce the general movement patterns of the entire population but do not consider the heterogeneity of individual attributes and movement patterns. Methods using machine learning are often capable of reproducing mobility data only for metropolitan areas where the training data were collected. Therefore, to generate nationwide human mobility data using this model, it is necessary to develop a cross-regional transferable learning method. In addition, because behavioral models are built using mobile phones, such as from GPS data, it is difficult to make the people movement data generated from such models available to other researchers and policymakers.

To overcome the inaccessibility issues of mobile phone datasets, large-scale people movement datasets for Japanese cities, such as PFLOW [14,15] and OpenPFLOW [16], have been developed and are open to researchers. PFLOW is generated by performing space-time interpolation processing on trip data collected from the Person Trip Survey (hereafter referred to as the PT Survey). As of April 2020, a dataset covering 25 locations in Japan (29 surveys) and 11 overseas locations is available to researchers, with it representing data on human flow for approximately 7 million people. An application is required to use PFLOW. OpenPFLOW is a more readily available dataset than PFLOW and was developed by combining open-data PT survey data in an OD aggregated format and national statistical data. Both datasets were based on the PT surveys. Therefore, the datasets could only cover cities in which the PT surveys were conducted. To achieve a fully accessible (open access) and comprehensive nationwide dataset of population movement, the transferability of open-data-generated human mobility models across different regions needs to be quantitatively tested and benchmarked.

To this end, this study developed the Pseudo-PFLOW, a pseudo-people-flow dataset that seamlessly covers the national scale. Pseudo-PFLOW was developed using only travel survey data collected in a limited metropolitan area and national statistical and spatial data available as open data. Pseudo-PFLOW represents the movement of people that is not a real movement of people but a pseudo-representation of typical daily behavior. Users of the data can identify trends in the urban spatial movement of people from Pseudo-PFLOW, and the dataset can also be used in a wide range of areas such as urban planning, commercial development, and simulation of infectious disease transmission.

- A method is proposed for generating pseudo-people-flow data that represent typical daily behavior from travel survey data conducted in a limited urban area and open statistical data.
- The developed Pseudo-PFLOW is pseudo-people-flow data that represents the synthetic movement of approximately 130 million people of the total population of Japan.

- The accuracy of Pseudo-PFLOW was verified and benchmarked using mobile phone location data and trip survey data from multiple metropolitan areas as the evaluation data.

The remainder of this paper is organized as follows: Section 2 describes the related studies. Section 3 describes the method for generating pseudo-people-flow data, and Section 4 presents the results of the pseudo-PFLOW accuracy experiment. Section 5 discusses the experimental results. Finally, Section 6 concludes the study.

## 2. Literature review

### 2.1 Existing person movement dataset

Herein, we summarize the existing people movement datasets. In the past, people movement data were collected through questionnaires such as travel surveys. But recently, as information technology has advanced and become widespread, people's movement data has been collected using mobile phones and various sensors. Table 1 shows representative people movement datasets organized based on existing research [17–19].

| Kinds of data | Dataset | Description |
|---|---|---|
| Trip survey data | PFLOW [14,15], OpenPFLOW [16] | 7 million persons in 36 metropolitan areas |
| | My daily travel survey [20] | More than 12,000 households in Chicago metropolitan area |
| National census | United States' counties, Italy's provinces, Hungary's regions [21], Japan regions [22] | Commuting trip data; national Census |
| | U.S. population migration data [23] | Migration flow, statistics of income division of the internal revenue service (IRS) in the US |
| Basestation data | D4D challenge dataset [24] | Call detail records (CDRs) data generated from Orange's mobile phone users in Ivory coast |
| | Telecommunications-SMS call, internet-MI [25] | CDRs data generated by the Telecom Italia cellular network over the city of Milano |
| GPS data -Volunteers -Social media -Taxi | GeoLife[26], mobile data challenge [27] | Volunteers' log in China and Switzerland |
| | Brightkite, Gowalla [28] | Location log shared by social media users |
| | T-Drive [29], TaxiNYC [30], UberNYC [31] | Taxi log in Chengdu, Beijing and New York City |
| Other data -Bike station -Subway station -Road sensor | BikeDC, BikeNYC [32], BikeChicago [33] | Trip data between bike station collected from the New York City, Washington D.C., and Chicago bike system |
| | METR-LA [34], Performance measurement system [35] | Traffic volume and speed in the road of Los Angeles and California |
| | SHMetro, HZMetro [36] | Inflow and outflow data in metro station in Shanghai and Hangzhou |

**Table 1. Variety of person movement data**

**Travel survey data:** Generally, such data are collected through questionnaires in print media or on websites. In Japan, national and local governments are the main implementers, and travel surveys have been conducted in major metropolitan areas approximately once every 10 years (65 metropolitan areas, 143 times in total as of April 2020). Examples outside Japan include a travel survey of households in northeastern Illinois. Travel surveys are conducted through questionnaires that collect detailed information about the trip, such as individual and household attributes, purpose of the trip, transportation mode, and departure time. However, the high cost of surveys makes it difficult to conduct them frequently over a wide area. Consequently, only travel survey data for a limited number of cities exist.

**National Statistics data:** The census also contains data on people's movements. Typical examples include commuting data. County-by-county commuter flow data in the U.S. and state-by-state commuter flow data in Italy are publicly available. The people movement data included in the census covers the entire population. However, only certain purposes, such as commuting and moving in and out, are covered.

**Base station data**: This is log data about communications between the mobile phone and base station, and it records the movement data of people with base station location accuracy. Typical examples are call detail records (CDRs) that record the calls and data communication of cell phone users. Although the location accuracy of base station data ranges from several tens of meters to several kilometers, these data contain information for all cell phone users and have an extremely high population coverage rate. On the other hand, raw data is generally not published by mobile carriers to the outside world because it is extremely sensitive information that contains personal information. Some data are anonymized and made publicly available, but they cover only a limited area.

**GPS data**: Movement data recorded by a GPS receiver on a mobile device. Compared to base station data, GPS data have a higher positional accuracy of several meters to several tens of meters. Several datasets store data collected from a small number of volunteers and location-based social media content. In recent years, several datasets based on GPS receivers installed in cabs that store vehicle movement trajectories and passenger trips have been published. However, the GPS dataset has a very low sample rate. Datasets collected from cabs only exist for a limited number of cities. Additionally, it is not possible to understand the movements of non-taxi users from the dataset. Some studies [10, 37] have used national-level GPS data collected from commercial app users, but it is difficult for many researchers to use such data.

**Other data**: In recent years, people movement data collected from bicycle systems has become publicly available. Typically, these data are published in the form of flow data that include the location of the origin and destination bike stations and the time of use. Although these are not people flow or trajectory data, there are datasets of traffic volumes collected using sensors installed on the road. Similar to roads, there are also datasets for railroads, an important transportation infrastructure, whose dataset contains the number of passengers boarding and alighting at each station.

Currently, various forms of human mobility data exist as described above. However, no dataset seamlessly covers the entire country and can capture people's daily activities at an individual level. Travel surveys are limited to major metropolitan areas and do not provide information on people's movement in rural areas or across metropolitan areas. In addition, few large-scale travel surveys have been conducted outside of Japan. National statistics data cover the entire national population, but the mobility data it contains are limited in purpose and do not capture people's daily movement behavior. The purpose of this survey is also limited and does not represent people's daily movement behavior. However, the base station and GPS datasets that have been the focus of much attention in recent years contain highly sensitive data and are generally not published by mobile phone carriers. In recent years, commercial person flow and grid population distribution data aggregated from mobile phone logs have become available. Replica [39] offers a service that analyzes people's movement behavior within cities from various perspectives based on a dataset collected from mobile phone carriers and third-party applications. However, the datasets and services are expensive. In addition, the fact that the data are aggregated and are not raw data representing the movement of individuals limits their scope of utilization and users using them.

## 2.2 Human mobility modeling

The people movement dataset represents the movements of people in the real-world. Prediction models built from datasets are used to understand a person's movement in an environment changed by future and urban development. As a representative model, a four-stage estimation method was used to predict traffic demand. The four-stage estimation method was a traffic demand forecasting method and the process was divided into four stages. The four-step estimation method estimated the number of trips between zones and could not predict individual movements. In addition, the four-step estimation method could not address individual attributes and did not include the concept of time.

Since the 1990s, activity-based models [40–46] have been studied as a new travel demand forecasting method. The activity-based model is based on the principle that travel demand is a derived demand for people's activities, and involves the generation and scheduling of daily activities for everyone, subject to spatial and temporal constraints. In general, each activity contains activity information such as purpose, location, and duration. There are two typical approaches: the utility-maximizing model and computational process model. Utility-maximizing models use a discrete choice model to determine the content of activities to maximize individual utility. Bowman and Ben-Akiva et al. [40, 41] used a nested logit model to model the choosing process of activity patterns, time, place, and transportation mode. The computational process model, also known as the rule-based model, determines activities based on context-dependent heuristics. ALBATROSS [44] sequentially performed the determining steps. A skeleton of obligatory activities such as work or school, is defined in advance, from which leisure-purpose activities are added to determine the day's activities. The decision process in each step uses CHAID [47] constructed from travel survey data. TASHA [46] randomly generated activities from an empirical probability distribution created from travel survey data. The activities were then scheduled using a heuristic approach.

Established activity-based models, such as ALBATROSS and TASHA, require large-scale travel survey data. Therefore, the cost of building the model is high. In recent years, many studies have been conducted on human mobility behavior models using GPS and base-station data. TimeGeo [2] extracted an individual's home, work, and other locations from the CDR data by analyzing the time spent and frequency. Then, an individual behavior model is constructed using a time-inhomogeneous Markov model with parameters for everyone. Pang et al. [10] proposed a method to build a behavioral model from GPS data that considers the spatiotemporal preferences of individuals using reinforcement learning. However, existing models assume the existence of people movement datasets for target cities. In other words, without a dataset for training, it is not possible to generate human flow data covering the entire country.

## 3. Methodology

### 3.1 Target specifications

Based on travel surveys conducted in a limited metropolitan area and national statistical and open spatial data, this study developed pseudo-people-flow data that satisfy the requirements listed as follows:

**Pseudo-PFLOW:**
- Cover the entire population (households and individuals) in Japan.
- Should be compatible with agent simulations that consider households and individuals.

- Represent typical daily activities such as commuting, shopping, etc.
- Constructed using only open and inexpensive data that exist on a national basis.
- Have similar results as national statistical surveys on an aggregate level.
- The evaluation results are published using alternative data (e.g., mobile phone location data).
- This will not be fully open data but will be available upon request for research purposes.

The pseudo-people-flow data cover all of Japan and reproduce a person's movements. For individuals in the dataset, attributes related to households and individuals estimated from national statistical surveys were added, assuming agent simulation. Only typical daily activities such as commuting to work, school, and shopping were represented, and non-routine activities such as sightseeing were not covered. The people flow volume aggregated from the pseudo-people data was consistent with the national statistical data. When releasing the pseudo-people-flow dataset, the evaluation results were included so that users could understand the characteristics and quality of the dataset and use it appropriately. Note that the prior application of PT surveys in Japan was required. Therefore, the pseudo-people-flow data were not completely open but were released as a dataset that can be used by researchers if they request it.

## 3.2 Framework

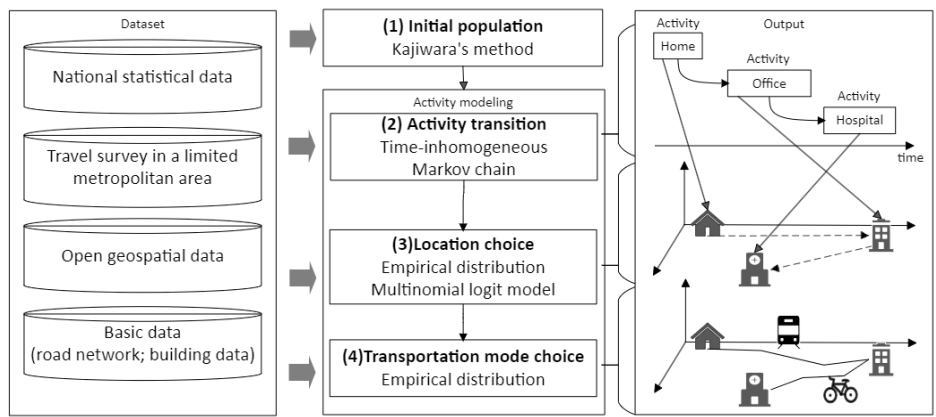

**Figure 1. Flowchart of Pseudo-PFLOW**

Figure 1 shows a flowchart for generating pseudo-human-flow data. In this study, an activity-based approach was used. In previous studies, models (2, 3, and 4 in Figure 1) were built based on the travel survey results. These models are strongly influenced by the city in which the travel survey was conducted, which makes it difficult to apply them directly to other cities nationwide. This study analyzed the content of activities that are common across cities from travel surveys in a limited metropolitan area and built models for only the common content from travel surveys. However, if national statistical surveys for the entire country were available, they were reflected in the activity content as much as possible. For example, the national census includes data on commuter and school trips. The statistical survey results covered the entire country and were highly reliable. These will also be conducted periodically in the future. Therefore, by adopting an approach that builds models based on national statistical surveys, it is possible to continually update the models, leading to the development of pseudo-human-flow data that reflects the survey year.

Table 2 lists the correspondence between each step in Figure 1 and the datasets. Commercial data on building location (No. 3 in Table 2) were also used in this study. In Japan, researchers can use No. 3 and No. 4 free of charge if they apply for a joint research administration system (JoRAS) with the Center for Spatial Information Sciences (CSIS) at the University of Tokyo [48]. It is also possible to substitute building data published in OpenStreetMap [49].

| No | Kinds of data | Dataset | Step in Fig. 1 |
|---|---|---|---|
| 1 | Number of households by attribute | National census (Open data) | (1) |
| 2 | Commuter OD (Origin-Destination) | | (3) |
| 3 | Building location | Zmap TownII (Commercial data) (Substitutable with OpenStreetMap) | (1) |
| 4 | Trip survey | Person trip survey in metropolitan area (Data published for scientific research) | (2) |
| 5 | Location of public facilities (Hospitals and schools) | Digital national land information (Open data) | (3) |
| 6 | Location of stations | | (4) |
| 7 | Elementary and junior high school districts | Public data of local government (Open data) | (3) |
| 8 | Number of employees and offices per 1km grid | Economic census (Open data) | (3) |
| 9 | Transportation choice ratio by attribute | National urban traffic properties investigation (Open data) | (4) |

**Table 2. Dataset for Pseudo-PFLOW**

## 3.3 Nighttime population with demographic information

Agent simulation requires population data that represents individuals with detailed attribute information such as age and gender. In this study, population data were generated using the household estimation method proposed by Kajiwara et al. [50, 51]. The method downscales national statistical data on households aggregated by subregion to the level of individual buildings and generates population data with the home location at the building level and attributes, such as age and gender (Figure 2).

This study assigned the categories of employed, young children, students, stay-at-home spouses/partners, and elderly individuals. In travel surveys and national statistical surveys, data are also aggregated into several categories to anonymize the data and facilitate the understanding of trends in the data. For example, in the labor force survey, the population was divided into working and non-working population categories; the working population was further divided into employed and unemployed; and the non-working population was further divided into commuters, stay-at-home spouses/partners, and others. This classification system is commonly used in national statistical surveys. In other words, by building behavior models for each of these categories, the granularity of the behavior model can be matched to national statistical surveys. The rules of role categorization are as follows:

- Employed: Probabilistically determined from the percentage of workers per municipality obtained from the labor force survey.
- Infants: Includes all children less than six years of age.
- Elementary and junior high school students: Elementary school students were defined as those between 6 and 12 years old, and junior high school students were defined as those between 12 and 15 years old.
- High school students and above: People who were 15 years of age or older. Probabilistically determined from the schooling rate for each municipality obtained from the basic school survey from those aged 15 years and older.
- Non-workers (stay-at-home spouses/partners; elderly): People other than workers and students. Elderly individuals were defined as those who were 65 years of age or older.

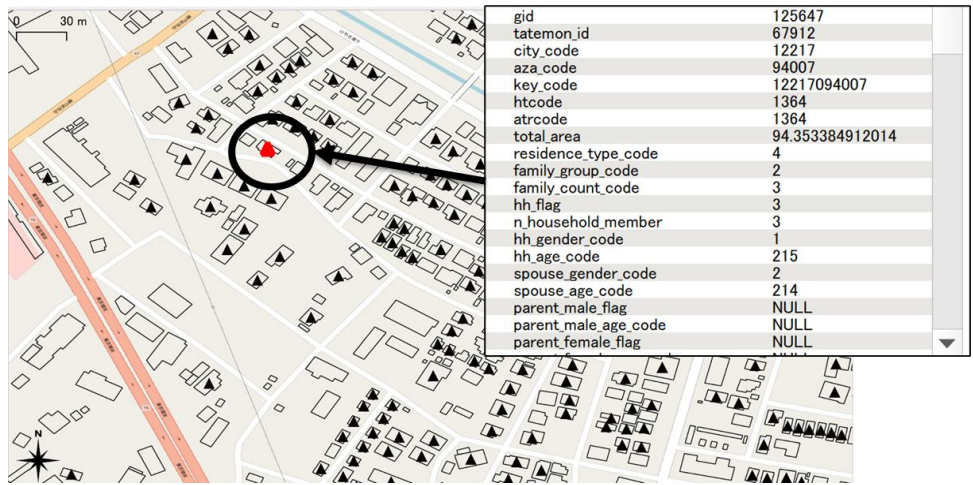

**Figure 2. Initial population data at the building level. Circle shows a household location**

### 3.4 Activity-based modeling

The relationship between the dataset and activity content is presented in Table 3. The first step (No. 2 in Figure 1) involved the generation of activities and schedules. In this process, the model built from travel surveys was applied nationwide to determine only the purpose, start, and end times of activities for the entire population. The next step (No. 3 in Figure 1) determined the location of the activity in a two-step process. After selecting a destination at the boundary level, a building was selected within the boundary. The last step (No. 4 in Figure 1) determined the transportation mode of the trip that was the derived demand for the activity. Details of each step are described in the following sections.

| Role type (Target age) / Activity element | Commuter (18 to 65 years old) | Elementary and junior high school students (6 to 15 years old) | High school and university students (16 to 22 years old) | Unemployed (All ages) |
|---|---|---|---|---|
| Activity transition (purpose + time) | Trip survey (No.4) | | | |
| Location — Boundary level | 1. Commuter OD (No. 2) 2. Number of employees per each grid (No.8) | Elementary and junior high school districts (No. 7) | Commuter OD (No. 2) | 1. Trip survey (No.4) 2. Number of employees per each grid (No.8) |
| Location — Building level | Building location (No. 3) | School location (No. 5) | School location (No. 5) | Building location (No. 3) |
| Transportation mode | Transportation choice ratio (No. 9) | | | |

**Table 3. Dataset for activity-based modeling**

The number in the table matches the number in Table 2.

#### 3.4.1 Activity transition

Trip-related data, such as activity locations and transportation modes, can be determined based on census data of commuters and school district data published by local governments. However, statistical data related to activity occurrence and schedules were not available. This study built a model based on travel survey data regarding activity generation, purpose, duration, and scheduling and generated national data based on this model.

In this study, we confirmed the validity of our approach by analyzing the differences between cities in terms of the content of activities modeled from travel surveys. This analysis covered the Tokyo, Kinki, and East Suruga metropolitan areas. The Tokyo metropolitan area is the largest metropolitan area in Japan with a population of approximately 36 million. The Kinki metropolitan area is a multinuclear metropolitan area with several coexisting core cities. The East Suruga metropolitan area is local.

First, a motif analysis [7] was performed to characterize people's movement patterns. Motif analysis represents human behavior using a simple network structure with visited locations as nodes and trips as directed edges between the nodes. The results of motif analysis are shown in Figure 3. The results in Figure 3 show the differences in travel patterns by people's roles; however, there are no differences by metropolitan area. Next, the percentage of people going out in different time zones was analyzed to determine the characteristics related to people's activity times (Figure 4). The results show that there is no difference in the percentage of people who leave metropolitan areas. Finally, the percentage of trips by purpose was analyzed, as shown in Figure 5. The results show that the difference by metropolitan area was very small, ranging from 1 to 2%.

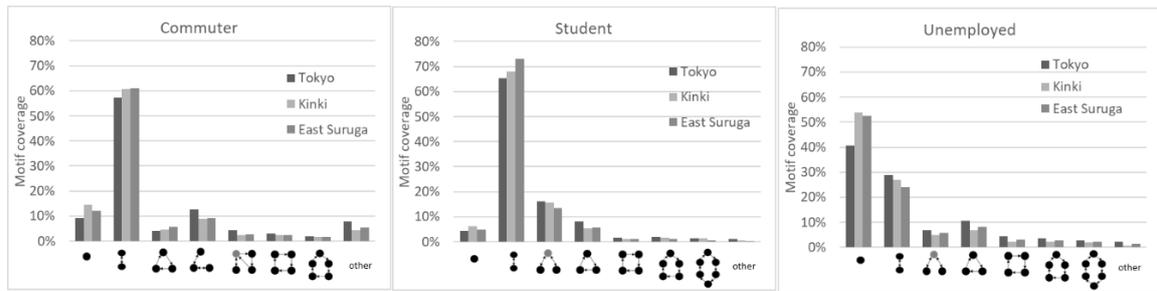

**Figure 3. Similarity of motif pattern ratio of each metropolitan area**

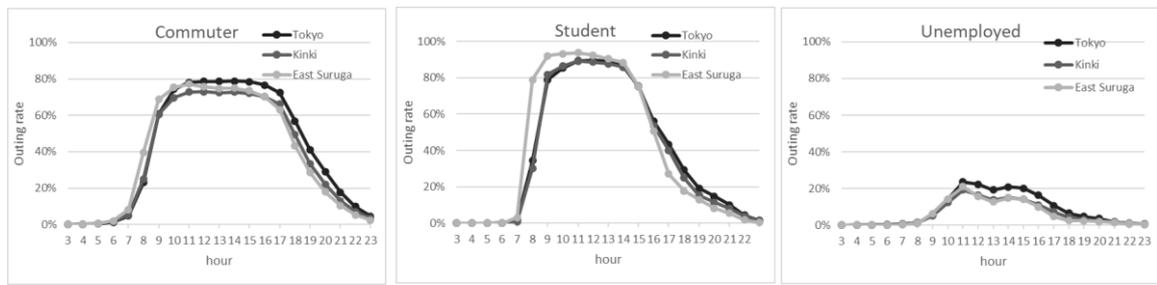

**Figure 4. Similarity of outing ratio of each metropolitan area**

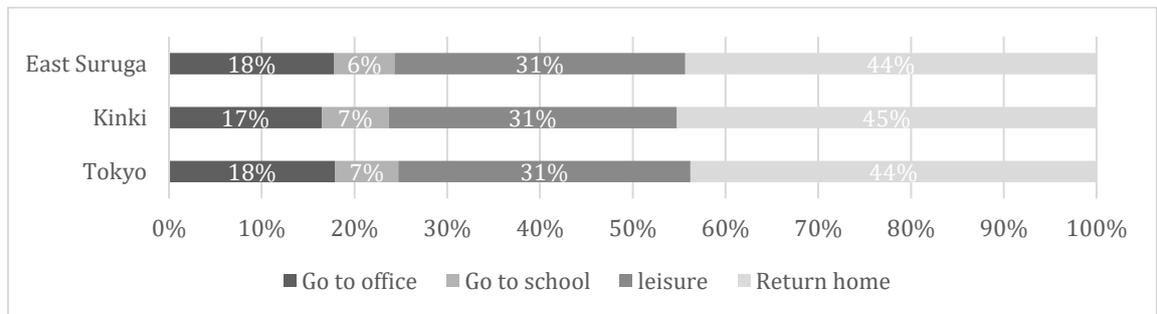

**Figure 5. Similarity of trip ratio by purpose**

Based on the results of the analysis, this study concluded that there were no significant differences in the occurrence, duration, and transition of activity among cities in Japan and generated nationwide person movement data using a model built from travel surveys from a single city. Specifically, we used a time-inhomogeneous Markov chain model [2]. $i$ is the state at time $k$, $j$ is the state at time $k + 1$, and the transition probability $p_k(i,j)$ from state $i$ to state $j$ can be expressed as shown in Equation 1.

$$p_k(i,j) = P\{X_{k+1} = j | X_k = i\}. \tag{1}$$

Markov states are defined for eight objectives: home, work, school, four leisure activities (shopping, eating, hospital visits, and others), and business. The transitions to different states represent the occurrence of new activities. Loop transitions in the same state represent a continuation of the activity duration. Transition probability matrices are determined separately for gender and elderly/non-elderly categories, in addition to personal roles. Therefore, the model reproduces differences in activity transitions based on personal attributes.

### 3.4.2 Location choice

The location of the activity was selected in two steps using statistical surveys and spatial data for each objective. First, the location was chosen at the boundary level, such as the city, grids, or school districts. Next, the location at the building level was selected using building data within the area selected in the previous step. The following section describes the location choice method used for this purpose:

- **Commuting**: In the first step, the destination city was selected probabilistically at the municipal level based on empirical distribution data of commuter trips from the national census of Japan. Then, using the number of employees obtained from the economic census as the allocation probability, a grid was selected within the destination city. If the destination and home were in the same city, the grid was selected using the Huff model [52], where the number of employees is the utility and the inverse square distance was the resistance because distance resistance has a significant impact. In the second step, a building was selected within the destination mesh using the area of the building as the distribution probability.

- **Commuting (Elementary and junior high school students)**: Using school district data and school location data published by the local government, the destination school was determined based on the location of homes.

- **Commuting (Students above high school)**: As a first step, the destination city was selected probabilistically based on the empirical distribution data of commuter trips included in the national census of Japan. Next, destination schools were randomly selected from within the selected city. If the number of students per school was known, it was used as the selection probability.

- **Leisure (shopping, hospital visits, etc.) and business:** No national statistical survey data exists for leisure trips. Therefore, the first step was to build a multinomial logit model constructed from existing travel surveys to select the activity location. The utility function of the model included the utility variables listed in Table 4. To address the differences in the spatial resolution between travel surveys, the resolution of the options was the city-level encompassing the travel survey zone. Additionally, to address differences in city size, the optimal parameters were calculated for each city size (less than 100,000; 100,000 to 500,000; and 500,000 or more). In the second step, as in the commuting process, a destination grid was selected within the city and then a destination building was selected within the destination grid. In the grid selection process, the allocation probabilities were determined using the number of employees for the business type corresponding to the trip purpose. For example, if the purpose was a hospital visit, the value of the number of employees in the medical field was used.

| Utility variable | Dataset No in Table 2 |
|---|---|
| Distance between cities | - |
| Dummy variable for gender | No. 1 |
| Dummy variable for senior (people over 65 years old) | No. 1 |
| Population density of the destination city | No. 1 |
| Business office density of the destination city | No. 8 |

**Table 4. Utility variable of location choice model**

### 3.5.3 Transportation choice

In Japan, apart from trip surveys conducted in metropolitan areas, a national cross-sectional urban transportation characteristics survey is conducted throughout the country. The survey provides information on transportation modal share for each of the following subcategories: metropolitan area scale and individual attributes such as gender, age, travel distance, and purpose. By mapping the survey results to each attribute, this present study probabilistically assigned transportation modes to trips between activities. If a rail was assigned to a trip, the trip was split into three sub-trips: origin to departure station (access trip); departure station to arrival station; and arrival station to destination (egress trip). Then, access and egress trips were also probabilistically assigned to the transportation mode, except for the training mode. As an exception, if there was no station within 3 km of the trip's departure and arrival points, the transportation mode was assigned to the candidates, excluding the train. In addition, the same transportation mode was assigned to commuting to and returning home.

## 4. Data evaluation

### 4.1 Generation of pseudo PFLOW

#### 4.1.1 Data processing

In this study, pseudo-people-flow data covering all of Japan was developed. First, based on the methodology described in Section 3, trip data were generated for the entire population of approximately 130 million people. In this experiment, a behavioral model constructed using only 2008 PT survey data for the Tokyo metropolitan area was used. Of course, it is more accurate to generate pseudo-people-flow data from PT surveys in multiple metropolitan areas, but we used PT survey data of other metropolitan areas as evaluation data for pseudo-people-flow data. Next, a spatiotemporal interpolation process [14] was performed for each trip to determine the position of each person. The spatiotemporal interpolation process is the process of route choice of the trip and determines the position at each defined interval of time. For trip route selection, Dijkstra's method was used to find the route with the shortest travel time. DRM data [53] were used for the road network, and the dataset developed by Kanasugi et al. [54] was used for the railway networks. By setting traffic speeds according to the link types, such as 100 km/h for highways and 30 km/h for general roads, routes could be selected based on the characteristics of the network. The speed of the rail link was set to 60 km/h.

In this study, the data utilization society creation platform MDX [55] was used as the computer resource. In detail, an instance with 100 virtual CPUs and 160 GB of memory was built on MDX and processed in parallel for approximately 10 days to generate pseudo-people-flow data for the entire country.

The visualization results of the pseudo-people-flow data are shown in Figures 6 and 7. Figure 6 shows the raw pseudo-people-flow data displayed as individual dots. Kepler.gl [57] was used for the visualization. We knew that the number of people in the center of Sapporo was small at 6 a.m., but many people gathered in the center of the city at 12 p.m. Figure 7 shows the visualization results of the population distribution and trip flow volume aggregated from the raw pseudo-people-flow data.

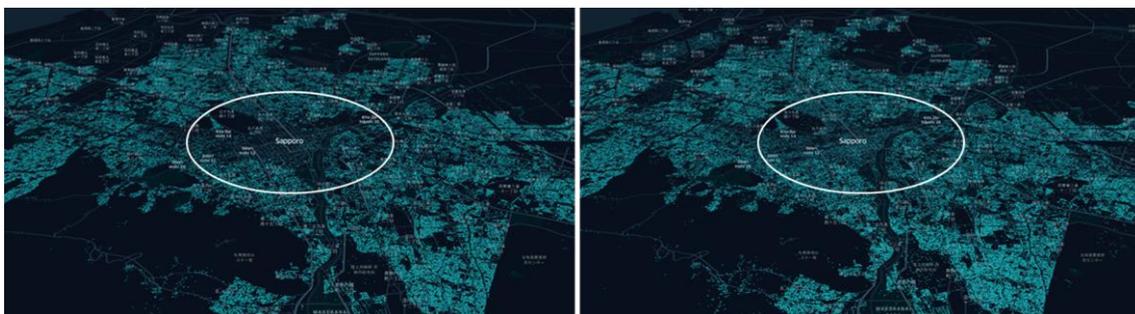

**Figure 6. Raw data of pseudo PFLOW. The figure shows the people's locations in Sapporo, Hokkaido, with a population of approximately 2 million. Left is the result at 6 a.m. Right is the result at 12 a.m. We can see the change in the distribution in the center of Sapporo.**

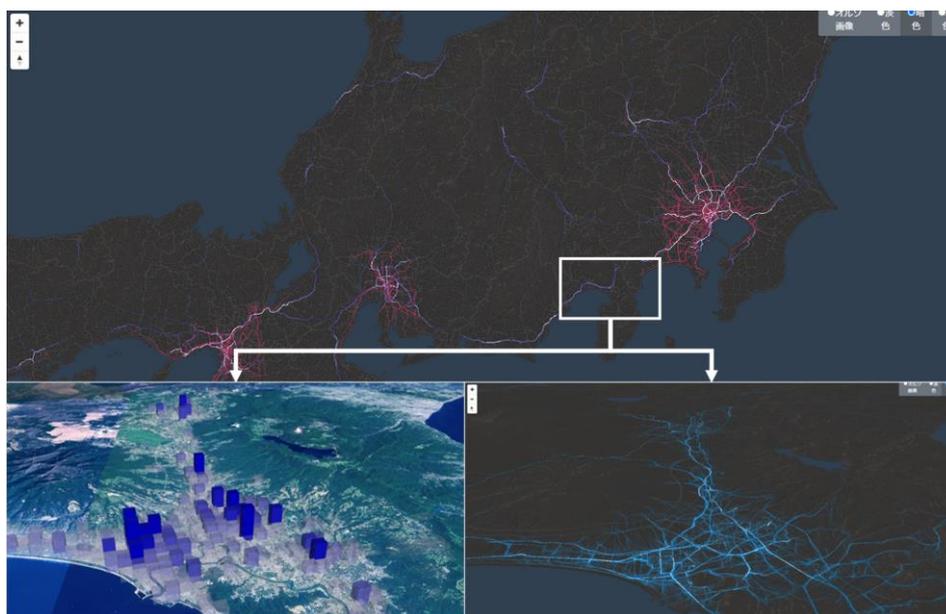

**Figure 7. Aggregated data of pseudo PFLOW. The figure above shows the link traffic volume in a wide area of Japan. The figure below shows the person flow data of the East Suruga metropolitan area. The left figure is the 1km grid population distribution and the right figure is the link traffic volume.**

### 4.1.2 Data publishing

One of the objectives of this study was to make person flow data available to researchers. In this study, pseudo PFLOW was made available to researchers worldwide through the Joint Research Access System (JoRAS) [48] of CSIS at the University of Tokyo. JoRAS is a data infrastructure developed to support and promote research related to spatial information science. Researchers can use the datasets maintained by JoRAS free of charge by submitting a joint research application with the CSIS. The PFLOW [14, 15], developed based on PT surveys, is also available to researchers through JoRAS. The dataset provided by JoRAS, including PT survey, road network, and building data, was used in the development of pseudo PFLOW.

The following data of Pseudo PFLOW are published on the web page (https://joras. csis. u-tokyo. ac. jp/dataset/list_all) of JoRAS and People Flow Project (https://pflow.csis.u-tokyo.ac.jp/data-service/pflow-data-2/). The public page provides raw, aggregate, and model data. The raw data represent an individual's trajectory. The aggregation data represent traffic volume aggregated by link and population distribution aggregated by 500×500 m$^2$ grids. The model data provided the parameters of the Markov chain used for the activity transitions and the Multinomial Logit Model to determine the activity location.

## 4.2 Evaluation of pseudo PFLOW

The accuracy of the pseudo-human-flow data was evaluated using three indices: population distribution, trip volume, and trip coverage. The population distribution was evaluated for Toyama and Shizuoka prefectures (green in Figure 8) because of the availability of cell phone data for them. The trip volume was evaluated for two metropolitan areas: the Kinki metropolitan area (blue in Figure 8), with a population of approximately 20 million among the large metropolitan areas, and the East Suruga metropolitan area (orange in Figure 8), with a population of approximately 650,000 from among the local metropolitan areas. Note that yellow area in Figure 8 represents the area of the Tokyo metropolitan area PT survey used to generate the behavior model for pseudo-people-flow data.

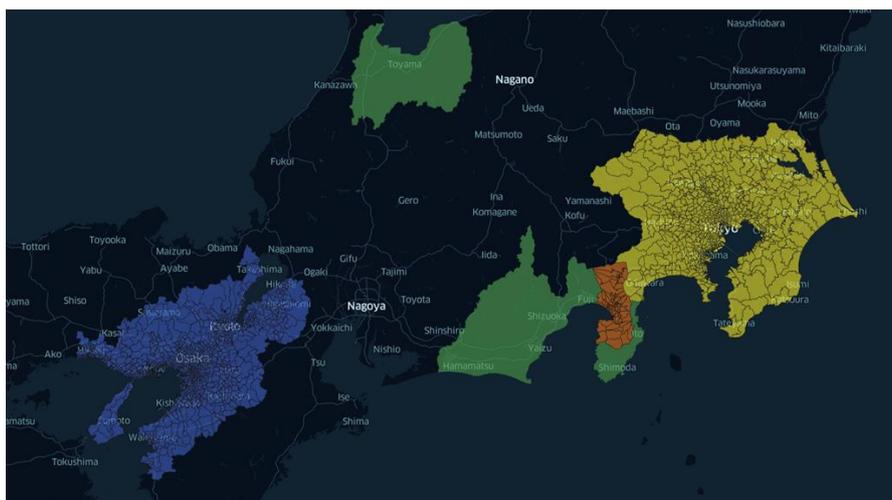

Figure 8. Target areas for evaluation. The green area has Shizuoka and Toyama prefectures that were the target areas for population distribution evaluation in 4.3.1. The yellow area is the Tokyo metropolitan area. The blue area is the Kinki metropolitan area. The orange area is the East Suruga metropolitan area. These were the target areas for trip volume evaluation in 4.3.2.

### 4.2.1 Population distribution

Using mobile phone data as evaluation data, the accuracy of the population distribution of pseudo-human-flow data was evaluated. The mobile data used were population distribution data on 500×500 m$^2$ grids estimated from the base station logs of DoCoMo, which had the largest market share in Japan at 36.9% as of the end of September 2020. The population distribution estimated from the mobile phone data was obtained by allocating the number of users within the coverage area of the base stations to which they were connected (several hundred kilometers) and then expanding the data so that the total number matched the population.

To confirm the accuracy of the mobile phone data, this study compared census and cell mobile data and census and pseudo-people-flow data for the population distribution at 6 a.m. The national census was conducted as a full-count survey because the data were used as the basis for national laws and administrative measures. Therefore, it maintained highly accurate night-time population values. Therefore, we compared the population distribution at 6:00 a.m., before many people started their activities with the national census.

The results are shown in Figure 9. The left side of Figure 9 shows the results of the comparison between the national census data and the pseudo-people-flow data. As mentioned in Section 3, the population data for the pseudo-people-flow data were generated from the national census, resulting in a very high R2 of 0.88. The right side of Figure 9 shows the results of the comparison between the census and mobile phone data, indicating that the R2 was generally highly accurate at 0.6. However, the error was not small compared with the pseudo-people-flow data. These results indicate that mobile phone data do not necessarily represent real values. However, it is the only data that can be used as evaluation data, as no alternative data exists.

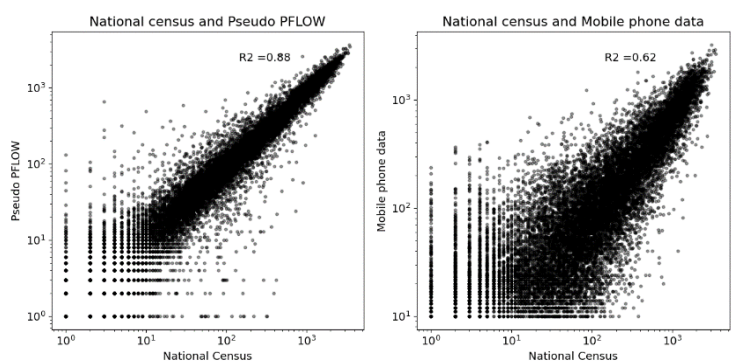

**Figure 9. Evaluation of initial population distributions in Shizuoka and Toyama prefectures. Left: Nation census and mobile phone data at 6 a.m. Right: National census and pseudo PFLOW at 6 a.m.**

The results of the comparison of pseudo-people-flow data with mobile phone data are shown in Figures 10 and 11. Figure 10 shows a scatter plot of the values at 12:00 p.m. to obtain a rough idea of the error. From left to right in Figure 10, the comparison was conducted at different accuracies: administrative boundaries (large scale), 1000×1000 m$^2$ grid (medium scale), and 500×500 m$^2$ grid (small scale). The results showed that the accuracy decreased as the scale decreased, but even the finest 500×500 m$^2$ grid showed generally high accuracy with an R2 of 0.61. However, the 500×500 m$^2$ grid has a large scatter of points on the scatterplot. As shown in Figure 9, the error in the real-world population distribution is likely to be larger when errors in the mobile data are considered. In other words, when evaluated rigorously, the accuracy of the 1000×1000 m$^2$ grid is considered the limit of use for pseudo-people-flow data.

Figure 11 shows graphs of R2 values by the time of day from 6:00 a.m. to 8:00 p.m. for each of the scales described in Figure 10. To confirm the effectiveness of the behavioral model, the R2 values are shown with mobile phone data for each time period and pseudo-people-flow data at 6:00 a.m. The results show no change in accuracy by the time of day at either scale. The evaluation at scales smaller than the administrative boundary level shows that the comparison results between the mobile phone data and pseudo-people-flow data at 6:00 a.m. is worse toward midday when many people are active. The results indicate that the behavioral model replicates the movement of people, capturing changes in population distribution.

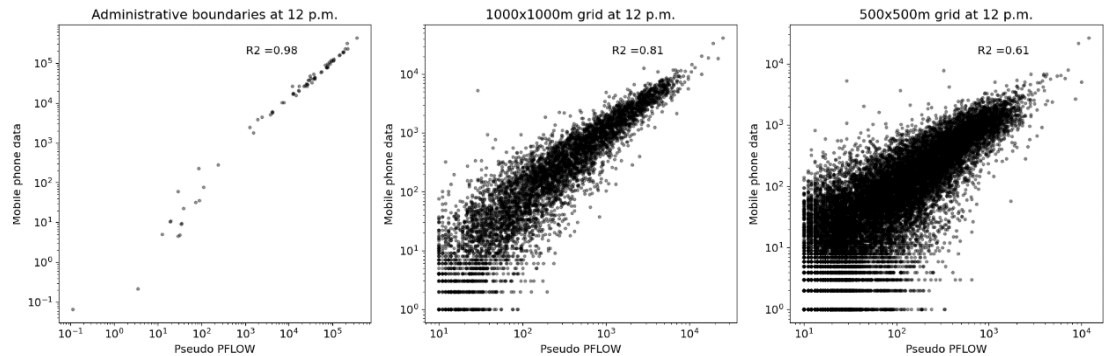

**Figure 10. Evaluation of population distribution for each area scale**

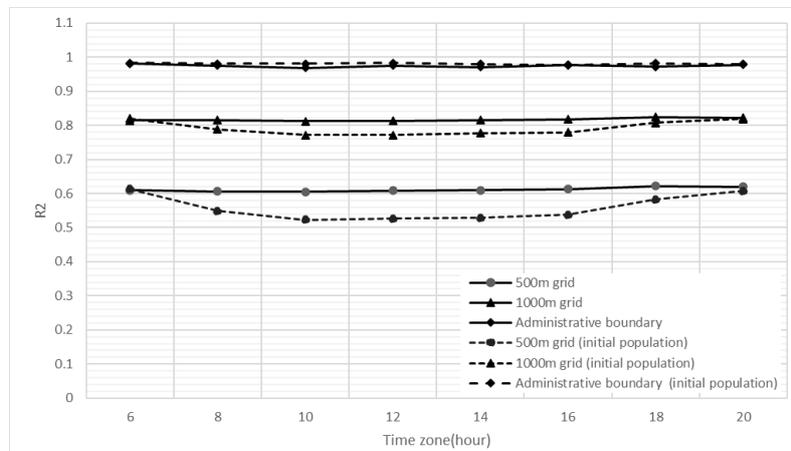

**Figure 11. Evaluation of hourly population distribution for each area scale**

### 4.2.2 Trip volume

In the evaluation data, trip volume was compared using the Tokyo metropolitan area PT used to generate the pseudo-people-flow data and the metropolitan area PT data not used to generate the pseudo-people-flow data. The results of the comparison of the daily trip volumes by purpose are shown in Figure 12. The Tokyo metropolitan area PT was only used to build a model to determine the location of leisure activities. The reason for including the Tokyo metropolitan area PTs in the evaluation data was to check the fitting of the model and use it as a basis for evaluation against other metropolitan areas' PT data.

The results of the commuting trip show that R2 is generally high in the Tokyo and Kinki metropolitan areas. As shown by the graph on the right, R2 is less than 0.5, which is not accurate for the East Suruga metropolitan area that is a local metropolitan area. Commute trips were generated from commute trip data from the national census. The commute trip data were aggregated by administrative boundaries, but the boundaries of rural areas were larger than those of urban areas. The low accuracy of spatial location estimation may be the cause of the difference in the R2 values.

Because school district data are publicly available, trips were generated for elementary and junior high school students with a high degree of accuracy. Unlike commuting, the same level of accuracy is achieved in the local metropolitan area of East Suruga as in the major metropolitan area. On the other hand, trips for high school students and above, as with commuting, are generated based on national census data and are therefore less accurate than those for elementary and junior high school students. Trips for high school students and those older than them are more diverse and smaller in volume than trips for elementary and junior high school students because they choose where to go to school based on their family circumstances and future goals. For the above reasons, the results in the middle panel of Figure 12 are more accurate than commuting for trip patterns with larger volumes but hardly different in accuracy from commuting for trip patterns with smaller volumes.

The evaluation results for leisure trips are generally highly accurate, with an R2 of 0.5 or higher in all metropolitan areas. Leisure trip locations were selected based on a multinomial logit model constructed from the Tokyo metropolitan area PT data. The evaluation results for the Kinki and East Suruga metropolitan areas show almost the same accuracy as that of the Tokyo metropolitan area. This result indicates that a model constructed from a PT survey of a limited metropolitan area could be applied nationwide.

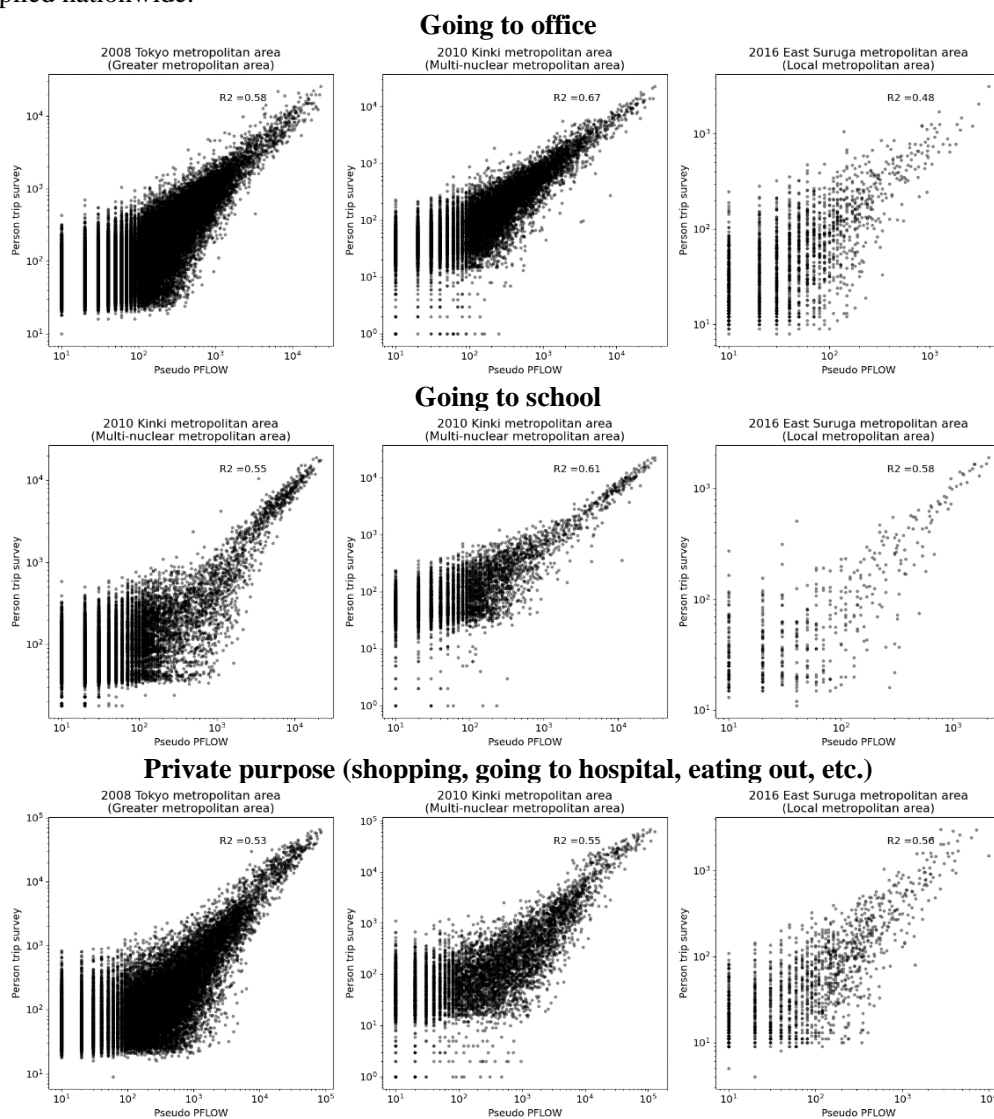

Figure 12. Evaluation of trip volume at PT zone level

### 4.2.3 Evaluation of trip coverage

Finally, the trip volume by purpose was compared in the metropolitan area and metropolitan area as evaluation data. The results of the comparison of trip volumes by purpose in each metropolitan area are shown in Table 5. The results for the Tokyo and Kinki metropolitan areas indicate that the error in trip volume for each purpose is approximately 10%. In the East Suruga metropolitan area, the volume of commuting trips is estimated to be 17% lower, and the effect of this is that the volume of home trips is insufficient. The generation amount of commuter trips is determined by the number of commuters obtained from the national census; therefore, the volume should be estimated with a high degree of comparative accuracy. We will collect other survey data for this case and proceed with a more detailed investigation of the cause in the future. Table 5 shows that an overall high coverage was achieved.

| Trip purpose | Tokyo metropolitan | Kinki metropolitan | East Suruga metropolitan |
|---|---|---|---|
| Going to office | 95% | 105% | 83% |
| Going to school | 104% | 113% | 102% |
| Leisure trip | 96% | 106% | 90% |
| Returning home | 93% | 97% | 84% |

**Table 5. Evaluation of trip coverage. The values show the value obtained by dividing daily pseudo trip volume by daily PT trip volume.**

## 5. Discussion

By analyzing the results of PT surveys in several metropolitan areas, as described in Chapter 3, this study found commonalities related to activity generation and transitions. A time-inhomogeneous Markov model was constructed regarding activity patterns and the occurrence, duration, and length of activities based on existing PT surveys and it was used to reproduce the movement of people across the entire country. Activity locations were determined by utilizing nationally surveyed statistical data that considers the characteristics of different cities. As a result, the pseudo-people-flow data generated by the proposed method can represent various patterns of people's movement. The evaluation results in Chapter 4 show that the pseudo-people-flow data achieve values close to those observed in the real world for population distribution and trip volume indicators. However, the current method uses only a portion of the existing statistical data and open data, and more elaborate people movement data are expected to be generated in the future by increasing the amount of data to be incorporated. In this study, a probabilistic selection method based on the building area of a business establishment was used to determine the location of the activity. In this process, data on the number of visitors per building can be used to determine the locations of activities that better reflect reality. In recent years, datasets such as the number of users of public facilities have been published by local governments in the context of open data initiatives. Therefore, we investigated and organized the available dataset for generating pseudo-people-flow data and considered how to incorporate them into the behavioral model appropriately, depending on the nature of the dataset.

In this study, pseudo-people-flow data were generated for Japan. In Japan, there are no large differences in socio-demographic contexts such as race and income. There were no large differences in the movement patterns of all the citizens. Therefore, as proposed in this study, the model constructed from PT survey data for limited metropolitan areas can be applied to the entire nation. However, the proposed method cannot be applied to countries with large differences in sociodemographic contexts. It is necessary to consider how to account for these contextual differences to generate pseudo-human-flow data from such countries. Moreover, in recent years, efforts to open various datasets such as urban planning, land use, population distribution, and transportation have been rapidly progressing worldwide [56]. Therefore, an environment for constructing pseudo-people-flow data is being developed not only in Japan but also globally. We look forward to the realization of an infrastructure for generating people's movement data from various alternative sources, not just from travel surveys and mobile phone data, and for the shared use of such data. We plan to organize the available data worldwide, and to what extent the method proposed in this study can be applied to other countries in the future. In the U.S. and Italy, as in Japan, commuter flow data are available; however, their spatial resolutions are different. In less-developed countries, official statistical data do not exist. Considering these circumstances, we will discuss the applicability of the proposed method and how to solve the problem if it is not applicable, in the future.

## 6. Conclusion

People movement data are fundamental for society and the use of people movement data is expected to produce results in a variety of fields. However, the protection of personal information and the high cost of obtaining such data are obstacles to the utilization of person movement data. In this study, we proposed a method for generating data representing pseudo-daily person movement by combining travel survey data from limited urban areas, statistical survey data, and open spatial data. Additionally, we developed pseudo-people-movement data for the entire population of Japan, consisting of approximately 130 million people, using the proposed method. The evaluation of pseudo-people-flow data was conducted using mobile phone data and travel survey data for metropolitan areas of different sizes. Population distribution and trip volume were evaluated, and generally, high accuracy was achieved with an R2 of 0.5 or higher for the comparison data.

Future research will improve the behavioral model to achieve the representation of diverse individual movement behaviors. By introducing microscopic traffic simulators and considering public transportation services, we can create a platform for the reproduction of realistic person movements in the simulation world.

## Acknowledgments

This research was the result of a CSIS joint research project (No. 1046) at the University of Tokyo. In addition, the data-processing platform MDX was used for people movement data generation. In addition, some of the results of this research were obtained under the JSPS Grant-in-Aid for Scientific Research 20K04718.

## Author contributions

**Takehiro Kashiyama**: Conceptualization, Data curation, Formal analysis, Investigation, Methodology, Project administration, Resources, Supervision, Validation, Visualization, Writing-review, and Editing.

**Yanbo Pang**: Investigation, Methodology and Validation.

**Yoshihide Sekimoto**: Supervision and Conceptualization.

**Takahiro Yabe**: Investigation, Writing-review, and Editing.

## Conflict of interest

There are no conflicts of interest to declare.